# Visual Pattern-Driven Exploration of Big Data

Michael Behrisch
*Harvard University*
Cambridge, USA
behrisch@g.harvard.edu

Robert Krüger
*Harvard University*
Cambridge, USA
krueger@g.harvard.edu

Fritz Lekschas
*Harvard University*
Cambridge, USA
lekschas@seas.harvard.edu

Tobias Schreck
*Graz University of Technology*
Graz, Austria
tobias.schreck@cgv.tugraz.at

Nils Gehlenborg
*Harvard Medical School*
Cambridge, USA
nils@hms.harvard.edu

Hanspeter Pfister
*Harvard University*
Cambridge, USA
pfister@g.harvard.edu

*Abstract*—Pattern extraction algorithms are enabling insights into the ever-growing amount of today's datasets by translating reoccurring data properties into compact representations. Yet, a practical problem arises: With increasing data volumes and complexity also the number of patterns increases, leaving the analyst with a vast result space. Current algorithmic and especially visualization approaches often fail to answer central overview questions essential for a comprehensive understanding of pattern distributions and support, their quality, and relevance to the analysis task. To address these challenges, we contribute a visual analytics pipeline targeted on the pattern-driven exploration of result spaces in a semi-automatic fashion. Specifically, we combine image feature analysis and unsupervised learning to partition the pattern space into interpretable, coherent chunks, which should given priority in a subsequent in-depth analysis. In our analysis scenarios, no ground-truth is given. Thus, we employ and evaluate novel quality metrics derived from the distance distributions of our image feature vectors and the derived cluster model to guide the feature selection process. We visualize our results interactively, allowing the user to drill down from overview to detail into the pattern space and demonstrate our techniques in a case study on biomedical genomic data.

*Index Terms*—Pattern Analysis, Pattern-Driven Exploration, Quality Metrics, Visual Analytics, User Guidance

## I. INTRODUCTION

Pattern analysis is becoming an increasingly important topic in many research domains ranging from natural sciences with subjects of physics, chemistry, and biology, to social sciences such as economics, public health, and sociology. Often, pattern analysis focuses on relationships between objects or phenomena. For example, in genomics, biomedical researchers study pairwise physical interactions between regions on the genomes [11]. They are interested in areas that are frequently in close contact, forming re-occurring structures [26], to understand the functional consequence of the spatial organization of the genome. Sociologists study connection and diffusion patterns in social networks [6] to understand information spreading and human relationships. In the transportation domain movements are analyzed to reveal frequent traffic patterns between cities and suburbs and optimize the infrastructure accordingly [54].

Thanks to nowadays's sensor, recording, and storage technology, scientists can rely on large datasets that may comprise valuable insights into their object of research. Hence, pattern mining in databases becomes more and more important but due to growing data sizes and complexity also increasingly challenging. Even when a suitable detection method is applied, patterns may be tough to interpret. Often hundreds or thousands of patterns are found, leaving it unclear which of them encode valuable information for the application domain, and how much support they have in the data.

To cope with these challenges, we contribute a stepwise analysis approach (see Figure 1), allowing to maintain overview in large and heterogeneous pattern-spaces. Given a set of found patterns and their visual representation we apply an image-based feature extraction to each pattern and cluster resulting vectors in a hierarchical manner. To evaluate the most suitable feature extractor, we calculate quality metrics on feature vector distribution and clustering results. The clustered pattern space can then be visually explored by interactively browsing the hierarchy, from overview to detail. While we believe the ideas of our approach to be generic, we focus on patterns in scatterplots and adjacency matrices, a common sub-problem in pattern mining.

The remainder of this paper is structured as follows: Section II discusses related work in the fields of feature extraction as well as automated and interactive visual pattern (-space) mining. We present the concept of our analysis pipeline in Section III, and subsequently outline our prototypical implementation in Section IV. After that, in Section V, we exemplarily evaluate our concept by instantiating the process in a case study on biomedical genomic data. We conclude by discussing limitations and future work in Section VI.

## II. RELATED WORK

Our approach combines (A) feature extraction methods to cluster frequently appearing visual patterns and enables (B) pattern space exploration using advanced navigation techniques. To evaluate feature extraction and clustering we adapt quality metrics (C). Hence, related work discussed in the following is threefold.

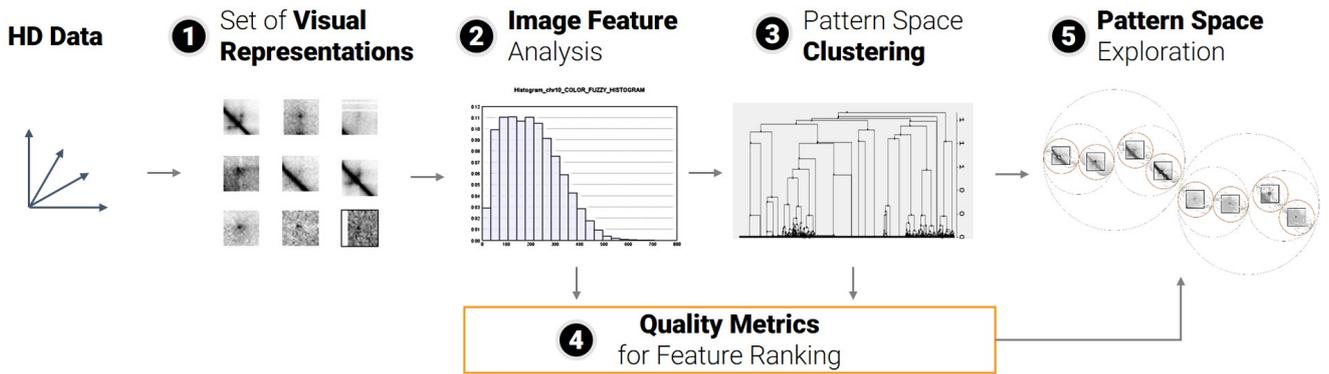

Fig. 1: In our pattern-driven exploration pipeline visual representations (1) serve as a proxy to understand the underlying data of interest. We extract image features (2) of these data views with the goal to partition the (visual) pattern space into interpretable, coherent chunks. To select an appropriate feature descriptor we calculate heuristic quality scores (4) assessing each descriptor's ability to discriminate the visualizations of interest. The clustering results (3) of the best performing feature descriptor, is shown in an interactive visualization that lets the user explore pattern clusters and -distributions (5).

*A. Image-Based Feature Extraction*

Image-based features allow characterizing data based on its visual representation. Commonly computed image-features comprise texture, color, and line/edge descriptors, shape, structure, and contour descriptors, interest point descriptors, as well as noise descriptors [38], [46]. As the extracted characterization are often similar to what the user visually inspects, they have been utilized to automatically suggest views of potential interest and to guide data exploration [10], [25], e.g., by feature-based aggregation and filtering. Doing so describes the data in a space that is different from the data itself and that instead is based on characteristics of visual patterns [10], [25], [40], [53].

Influential for this field is the work of Tukey [51] who formulates the problem that—as the number of plots to interactively inspect increase—exploratory data analysis becomes difficult and time-consuming. Tukey proposes to find the "interesting" plots automatically and to investigate those first. To that end, Wilkinson et al. [53] present a set of 14 measures for the quantification of distribution of points in scatter plots, called Scagnostics. Each measure describes a different characteristic of the data and helps, for example, to filter views with different Scagnostics measures than the majority. The underlying scatter plots are likely to exhibit informative relations between the two data dimensions.

Similar to these existing approaches we extract image-based features. By contrast, we apply the extraction not on visualizations showing the whole data but on snippets containing detected patterns and use the resulting feature vectors as an input to hierarchically cluster the pattern space.

*B. Semi-Automated Exploration*

As datasets grow in size and complexity, analysts run the risk of overlooking interesting patterns if relying on manual exploration only. To this end, intelligent methods for compressing and filtering data for potential patterns of interest have recently become a research focus in the visual analytics community.

Overview-based approaches aim to generate effective layouts over many candidate data portions, to efficiently spot patterns of interest. Examples include the Value-and-Relation display [55], which lays out pixel-oriented views based on their data similarity and allows for further drill-down. Another similarity-based layout is proposed by Ward and Guo [52], where many time series are represented by small glyphs.

Besides overview approaches, automatic filtering of views for potential structures of interest has been proposed. As mentioned in the preceding section, the Scagnostics approach [53] automatically analyzes structures in scatter plots, which can be used to rank and filter. In case class information is given, scatter plots can be filtered for discriminative views by class consistency measures [45]. Also, projection pursuit approaches, such as initially presented by Friedman and Tukey [18], try to identify interesting 2D subspaces in high-dimensional data (mostly depicted by scatter plot views). Further heuristic interestingness filters for Scatter- and Parallel Coordinate plots have been discussed [10], [49] and may narrow down the potentially large search space for high-dimensional data. While most approaches focus on global features, Shao et al. [43] propose means for determining frequent local scatter plot characteristics and summarize them in a motif-based dictionary.

Combing overview and filtering techniques, Tatu et al. [50] proposed an approach to analyze subspaces contained in high-dimensional data using hierarchical clustering for exploration on different levels of magnitude. Another approach, called ScagExplorer [9] clusters data based on Scagnostics [53] and allows for further drill-down and filtering. Also relying on clustering, Bach et al. [2] present an approach to aggregate adjacency matrices, so-called piles, indicating topological states in brain connectivity and to explore them at different levels of magnitude.

Similarly, our approach extracts features from patterns in matrices and clusters them to explore the result space. A main contrast to the work by Bach et al. is that our approach utilizes and evaluates image-based features to characterize and cluster the data and does not rely on a temporal order.

*C. Clustering Comparison*

The quality of clustering results depends on a dataset's characteristics and distributions but also on multiple consecutive data processing steps such as extracted features, distance measures, and the clustering strategy. In such a pipeline interim results add up and can lead to significant different clusterings. Researchers thus developed several statistical methods to evaluate and compare clustering quality, as summarized by Rodriguez et al. [35]. One widely used metric is the silhouette coefficient [37] comparing intra cluster cohesion to inter cluster separation. Another class of clustering comparison approaches, e.g., often used in the bioinformatics domain, make use of labeled reference data, where cluster memberships are already known [19], or they rely on domain knowledge to judge the quality of the clustering [14].

Our approach considers situations where no ground truth is given. Similar to measures summarized by Rodriguez et al. [35], we apply quality metrics to rank clustering results. However, instead of calculating metrics for one partitioning, we measure and compare the quality of the whole clustering hierarchy by recursively computing coefficients for all levels.

## III. Visual Pattern-driven Exploration Concept

Our goal is to analyze large amounts of high-dimensional data in a pattern-driven fashion. While the collected data may be rich in information, the exploration challenge is to find data decompositions, e.g., subspaces or data intervals, which expose the underlying core dataset characteristics. Finding and understanding these *central patterns* requires not only automated data analysis techniques but also the analysts' understanding, background knowledge, and experience. Figure 1 depicts our conceptual pipeline for supporting analysts in a *pattern-driven exploration* of large high-dimensional (HD) datasets.

*(1) Set of Representations*

Large-scale HD data analysis can hardly be accomplished on the raw data objects, and visual representation can help to facilitate a mental model. Hence, we center our analysis pipeline on the basic idea to regard the visual representations as a *proxy* to the data of interest, and base similarity and relevance computation tasks on the visual data representation, instead of the original (raw) data. Our aim in doing so is to provide user-friendly, interpretable and interactive assessment functions as a basis for search and analysis tasks.

In this work, we do not focus on the question *"Which visual representation is the best for the underlying dataset?"*, but rather start from the hypothesis that a data representation is established in the current analysis domain. In many domains, such as in the biomedical sector where results are frequently represented in heatmap displays [13], [23], [26], [27], or the environmental sciences, which oftentimes rely on scatterplot or line charts [7], [42], [44], analysts are oftentimes trained over years to see and retrieve patterns in a particular visualization type. While this assumption is not generally pertinent for all scientific domains, we claim that our exploration pipeline (Figure 1) is still applicable for the general task of understanding distributions in pattern spaces for a given data representation.

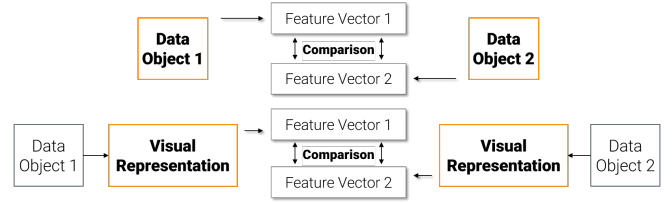

Fig. 2: Image Feature Analysis. Top: The standard feature extraction operates on the raw data and is typically defined in a static and heuristic way. Bottom: Our approach extracts features from a visual representation. The approach is able to visually represent why objects are similar and provides a starting point for navigation and visual pattern space exploration.

*(2) Image Feature Analysis*

Searching and analyzing are key tasks for retrieving, relating and reusing complex data sets. The standard approach to similarity computation extracts feature representations from the raw data (see Figure 2 (Top)). We propose to extract feature representations from the visual transformation of input data (Figure 2 (Bottom)). This approach may provide several advantages. First, a visual transformation of data is naturally linked with the user interface: visualization of the data is used in many applications, and it can be intuitively shown why two data representations are considered similar, e.g., by representing corresponding visual features in the visual data representation. A second main advantage is that one can implement visual analysis interfaces that allow the user to explore visual similarity of many data instances. We will develop this idea further in the steps *(3) Clustering* and *(4) Quality Metrics for Feature Ranking*. Finally, the similarity notion can flexibly adapt to user needs: different visual abstractions give rise to different similarity notions and feature extraction approaches and is thus a flexible means to adapt the exploration pipeline to the context of the user task.

One of the cornerstones of our pipeline is to employ image-based feature descriptors into the process. Many feature descriptors were proposed to automatically extract the most descriptive feature set from a given image or visual representation, among them novel visual descriptors that try to mimic human perception aspects. Applied in our context, these so-called *-gnostics approaches [3], [10], [25], [40], [53] can help the user to improve the interactive query specification and result interpretation stages of the visual analysis process.

*(3) Pattern Space Clustering*

Cluster-based navigation systems divide the exploration space into a range of distinctive clusters, such that each grouping corresponds to a meaningful data sub-selection. In the case of visual pattern-driven exploration, a "good" feature encoding applied on a clustering approach will result in groupings that reflect pattern-cluster memberships.

In this work, the central analysis task is to partition the pattern space into interpretable, coherent chunks. These chunks should expose similar visual patterns and guide the analysts to beneficial exploration paths for a subsequent in-depth analysis. Hierarchical or density-based cluster approaches [36] are consequently the data analysis method of choice since it allows the analysts to perceive (dis-)similarity of patterns, assess the pattern-to-noise ratio, and the pattern-distribution without predetermining a k-fold pattern space partitioning as done by, e.g., k-means clustering [29].

*(4) Quality Metrics for Feature Ranking*

The extraction of relevant information from high-dimensional data, is complex and time consuming. In that respect the notion *curse of dimensionality* represents a whole set of issues encountered in the analysis of these data sets: finding relevant data attributes, selecting meaningful and descriptive dimensions, removing noise represent are just a few of them.

Researchers have been trying to solve the aforementioned analysis problems through either automatic data analysis or interactive visualization approaches. However, we claim that integrated visual analytics approaches, where a machine searches automatically through a large number of potentially interesting data transformations and mappings based on quality metrics, and the user interactively steers the process and explores the output through visualizations will outperform isolated approaches.

Our work represents a specific example for the aforementioned quality metrics-driven exploration. We rely on a "good" image-feature extraction algorithm to derive an interpretable and useful pattern space clustering. One central assumption of our pipeline is that at least one image feature descriptor is capable to describe visual patterns. Earlier research has validated this hypothesis on relational data [3].

*(5) Pattern Space Exploration*

As mentioned before pattern space exploration is a human-centered analysis approach in which the automatic analysis should support critical parameters, i.e., the feature descriptor selection. On the other hand, finding an appropriate partitioning of the pattern space is not enough. These approaches need to involve an analyst who can make sense of the results and give them meaning in their respective analysis domain.

A visual interface for navigating in stratified pattern space clusters is hence necessary. Ideally, such a system provides Overview+Detail functionality, i.e., it allows to perceive pattern distributions, feature descriptor, and cluster uncertainty, and allows an interactive exploration of the pattern space for unexpected findings. We showcase an exemplary interactive exploration prototype in this work as depicted in Figure 4.

IV. PROTOTYPICAL IMPLEMENTATION

We found that a range of implementation challenges need to be tackled to instantiate our analysis pipeline presented in Section III. The findings that we derived during the implementation of our prototype will structure this section and outline potentially beneficial research directions.

*A. Quality Metrics Implementation*

One general challenge of pattern space exploration is that no ground truth data is available, making standard external evaluation metrics, such as precision or recall, inapplicable. Also, the exploratory nature of our analysis lets the user develop experience-based heuristics earliest after getting intermediate results. In order to bootstrap the analysis for retrieving interpretable—intermediate—results we developed quality metrics to rank our set of 26 feature descriptors for their ability to (1) *differentiate visual patterns* and to (2) *produce (visually) coherent groupings* of patterns.

*1) Pattern Discriminability:* We approximate a feature descriptor's ability to differentiate visual patterns by calculating statistics on the normalized Euclidean distance scores:

$$\sigma^2(\mathrm{FD}) = \frac{1}{\mathrm{n}^2} \sum_{1 \leq i \leq \mathrm{n}} \sum_{1 \leq j \leq \mathrm{n}} (ndist(\mathrm{FD(i)}, \mathrm{FD(j)}) - \bar{x})^2 \quad (1)$$

where $\mathrm{FD(i)}$ calculates the feature vector of the $i^{th}$ image in the dataset, $ndist()$ represents a normalized Euclidean distance and $\bar{x}$ corresponds to the average of all distance combinations. This variance calculation is a coarse-grained heuristic that solely allows the following conclusion: If there is a low distance score variance between the data feature vectors, the respective feature descriptor is not able to differentiate the inherent (visual) features; given the assumption that there are actually (human-)discriminable visual patterns.

Pattern discriminability for image feature descriptors has been initially studied, for example, in [3] in the context of matrix pattern research. However, more future research could be devoted for developing more fine-grained approximations. As an example, a correlation analysis of the image feature dimensions would be computationally more expensive but might result in potentially better retrieval of dimension subspaces that help in this context.

*2) Clustering Structure Quality Metric:* Pattern discriminability enables us to reject inappropriate feature descriptors early on, but it does not allow us to assess how well a feature descriptor can partition the feature space. We derive, therefore, a hierarchical clustering based on this feature descriptor and quantify its cluster separation and cohesion. A range of external measures quantifying the "accuracy" were presented in the past, such as the Rand index [32], the Fowlkes-Mallows index [16], and the Jaccard index [22], [30], which are only applicable to labeled datasets. In our case, however, we have to rely on internal quality measures that base their quality understanding on (dis-)similarities of feature vectors.

While also here a range of quality measures exist, i.e., the Dunn index [12] or the Calinski-Harabasz score [8], we chose silhouette coefficient [21], because it is represents an established and intuitive measure for both cohesion and separation of clusters. In our evaluation section (Section V-A) we present a comparative evaluation of the quality computations based on the Dunn index and the Calinski-Harabasz score. For our scenarios we found that the silhouette coefficient is well-suited. Its silhouette score bases on the mean intra-cluster distance and the mean nearest-cluster distance for each item in the dataset. A silhouette coefficient $SC$ for a partition with $k$ clusters is calculated by averaging the $k$ individual silhouette scores:

$$SC = \frac{1}{n} \sum_{1 \leq i \leq n} \frac{b(FD(i)) - a(FD(i))}{max(a(FD(i)), b(FD(i)))} \qquad (2)$$

where $a(FD(i))$ is the average dissimilarity between $FD(i)$ and all other points in the same cluster and $b(FD(i))$ is the average dissimilarity between $FD(i)$ and the data points in the nearest neighbor cluster. Since the number of clusters $k$ is unknown in our analysis we introduce a cut-level balanced silhouette coefficient with the following formula:

$$\phi(SC) = \sum_{1 \leq k \leq h} \frac{1}{2^{(k-1)}} \times SC_k \qquad (3)$$

where $h$ is the height of the clustering dendrogram tree. The score calculates for all hierarchical clustering cut-levels the silhouette coefficients and aggregates them with a weighting score depending on the cut-level depth. The intuition for this calculation is the following: A clustering dendrogram that shows many well-separated, but highly coherent, clusters on the higher levels should be favored over degenerated trees that show coherence, but little separation. Since our score aggregates over all clustering levels and most dendrograms contain long, degenerated sub-trees we weight their influence depending on the cut-level depth.

Our quality score bases on considerations about hierarchical clustering results. While this quality metric could potentially also be used for density-based clustering methods, future research and experiments should be devoted to proof it's generalizable.

*3) Compound Quality Score:* Both scores above are equally weighted and aggregated to derive a final quality score.

$$QM(FD) = 0.5 \times \sigma^2(FD) + 0.5 \times \phi(SC) \qquad (4)$$

In a future prototype, we are planning to let the user decide on the weighting and composition of the factors. With such a flexible understanding of quality, analysts could express—for example—their preference in scenarios where quality is sacrificed for faster computation times.

*B. Understanding Clustering Results*

In order to be of practical use cluster results need to be represented by prototypes that aggregate the information of its underlying cluster. While many approaches try to give an example-based cluster prototype, such as most medoid entity [31], other approaches try to aggregate the cluster information, as done by Strobelt et al. [48]. For our purpose to understand clustering results on visualizations exposing structured visual patterns, we developed an image-based visual aggregation for visual views, as Figure 3 depicts.

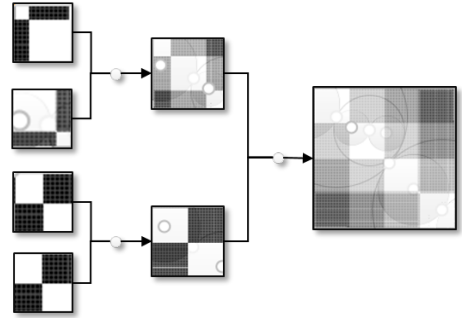

Fig. 3: Visual depiction of a cluster prototype in a hierarchical clustering. We overlay all cluster entities visually, such that their relative opaqueness value aggregates to 1.

As Figure 3 depicts, we construct a cluster representative from all entities contained in the cluster. We visually overlay each matrix image within one cluster such that their relative opaqueness value sums up to 1 (fully opaque). This opaqueness consideration shows the (un-)certainty of a hierarchical clustering with respect to the cut-level intuitively.

In our future research, we plan to investigate further visual aggregation methods, that focus on exposing the (un-)certainty of a hierarchical clustering. Since not all leaves are on the same height in a hierarchical clustering, one specific improvement will be to incorporate a leaf's path length to the root as a weighting factor.

*Navigation in Stratified Pattern Spaces.*

In this work, we showcase an interactive exploration system for pattern space exploration as depicted in Figure 4 on scatter plots. It follows the previously proposed framework for visual pattern exploration. Its purpose is to give an overview of pattern space clustering (Figure 4-1) and to enable the analysts to drill-down into the hierarchical—or stratified—cluster structure (Figure 4-2) to make sense of their certainty, quality, and cluster's membership composition.

As a proof of concept, we developed a hierarchical clustering visualization for general visualizations as an alternative to the standard dendrogram visualization for hierarchical clustering In our experiments we found that employing a radial layout is more space-efficient, thus allowing to render larger cluster prototypes. The user can switch between the two layout modes. In the radial layout, the cluster rings' sizes and

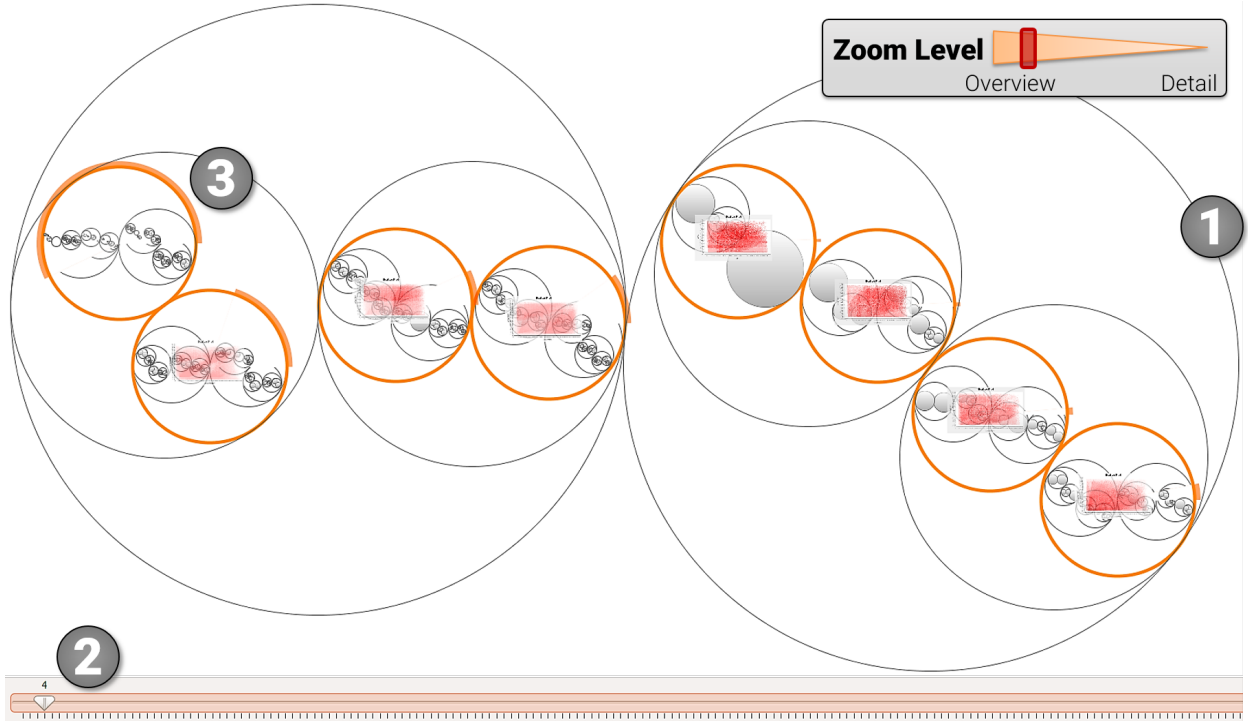

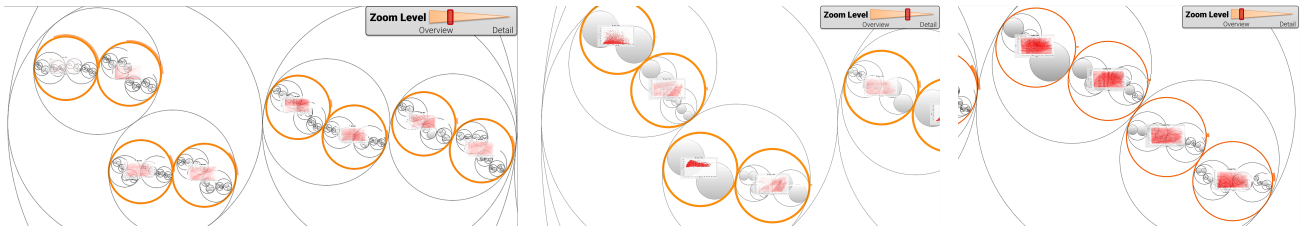

(a) Overview of the scatter plot pattern space ①; Dendrogram Cutlevel (DC) 4: The left cluster contains significantly more locally dense elements (1,200) than the right cluster with globally dense scatter plots (76). The drill-down slider ② allows modifying the dendrogram cutlevels and controls the cluster prototype resolution. The pie chart metaphor in ③ depicts the absolute cluster size.

(b) Left cluster DC5: scatter plot are subdivided by their highest density quadrants

(c) Left cluster DC8: Local monotonic gradients (flame pattern) becomes visible

(d) Right cluster DC4: Contains only high density scatter plots.

Fig. 4: Pattern space visualization for 1,276 scatter plots representing a BSRN earth observation dataset; The best performing feature descriptor COLOR_OPPONENT_HISTOGRAM differentiates globally dense from locally dense scatter plots on the first split levels and later differentiates the density aspects with its . The adapted dendrogram visualization shows a cluster representative image and encodes the size of the respective cluster in the outer arcs. The orange border highlights the current dendrogram split level, which can be interactively modified with the slider below.

positions reflect the hierarchy. The cluster's membership size is depicted additionally in the ring's outer appearance: a pie chart metaphor Figure 4-3 shows the percentage of members in this cluster with respect to the overall number of data items; Small clusters (small arc angle) will be distinguishable from larger clusters (large arc angle). We render the cluster prototypes into the ring's center, such that all leaf nodes contribute equally to the prototype. As Figure 4-2 depicts a "drill-down slider" can be used to examine a cluster's membership composition in the form of cut-levels in a dendrogram. This interaction allows retrieving meaningful dendrogram split levels and thus an appropriate number of visual pattern clusters in the dataset.

We found that our implemented interactive panning and zooming operations are helpful but may become tedious. On an abstract level, however, these interactions allow us to derive implicit and explicit interestingness considerations from the visualized pattern space. In a future work, we want to make use of these aspects by incorporating relevance feedback and learning, following the line of research suggested, e.g., in [4].

## V. EVALUATION

We evaluate our pipeline implementation with two distinct experiments. First, we focus on the parameter choices, outlined Section IV-A, for constructing our quality metrics, second, we present in Section V-B two case studies on (a) earth observation data and (b) on genome interaction data (Section V-C) to showcase the applicability of our approach on a real-world dataset.

| FeatureDescriptor | Time_AVG | D_Variance | D_Pop Var | D_STD | C_Silhouette Index | C_DunnIndex | C_Calinski Harabasz | Q_ScoreSilhouette | Q_ScoreDunn | Q_ScoreCalinskiHarabasz |
|---|---|---|---|---|---|---|---|---|---|---|
| COLOR_OPPONENT_HISTOGRAM | 13.75 | 0.04 | 0.04 | 0.20 | 1.46 | 0.17 | 10,593.26 | 0.75 | 0.10 | 5,296.65 |
| COLOR_FUZZY_OPPONENT_HISTOGRAM | 56.64 | 0.04 | 0.04 | 0.19 | 1.44 | 0.05 | 3,737.96 | 0.74 | 0.04 | 1,869.00 |
| COLOR_AUTO_COLOR_CORRELOGRAM | 29.94 | 0.04 | 0.04 | 0.19 | 1.40 | 0.03 | 4,608.85 | 0.72 | 0.04 | 2,304.44 |
| TEX_GRADIENT | 234.13 | 0.04 | 0.04 | 0.19 | 1.38 | 0.01 | 9,395.36 | 0.71 | 0.02 | 4,697.70 |
| TEX_HARALICK | 13.34 | 0.06 | 0.06 | 0.24 | 1.36 | 0.23 | 7,077.16 | 0.71 | 0.14 | 3,538.61 |
| COLOR_GLOBALCOLORHISTOGRAM | 20.36 | 0.04 | 0.04 | 0.20 | 1.33 | 0.04 | 7,208.42 | 0.69 | 0.04 | 3,604.23 |
| POI_FAST | 256.00 | 0.03 | 0.03 | 0.18 | 1.33 | 0.03 | 4,877.54 | 0.68 | 0.03 | 2,438.79 |
| POI_SURF | 64.01 | 0.02 | 0.02 | 0.13 | 1.34 | 0.06 | 1,681.51 | 0.68 | 0.04 | 840.77 |
| COLOR_FUZZY_HISTOGRAM | 641.59 | 0.03 | 0.03 | 0.18 | 1.31 | 0.11 | 4,049.75 | 0.67 | 0.07 | 2,024.89 |
| STRUCTURE_JPEGCOEFFICIENTHISTOGRAM | 203.27 | 0.03 | 0.03 | 0.17 | 1.26 | 0.13 | 1,336.61 | 0.64 | 0.08 | 668.32 |
| TEX_LOCAL_BINARY_PATTERN | 129.05 | 0.03 | 0.03 | 0.18 | 1.24 | 0.04 | 3,549.06 | 0.64 | 0.04 | 1,774.55 |
| TEX_TAMURA | 59.83 | 0.03 | 0.03 | 0.18 | 1.24 | 0.11 | 4,582.44 | 0.63 | 0.07 | 2,291.24 |
| TEX_GABOR | 21.38 | 0.03 | 0.03 | 0.16 | 1.21 | 0.01 | 5,004.97 | 0.62 | 0.02 | 2,502.50 |
| NOISE_STATISTICALSLIDINGWINDOW | 37.55 | 0.02 | 0.02 | 0.15 | 1.21 | 0.01 | 5,424.84 | 0.62 | 0.02 | 2,712.43 |
| SHAPE_COMPACTNESS | 5.44 | 0.02 | 0.02 | 0.13 | 1.21 | 0.02 | 4,112.12 | 0.61 | 0.02 | 2,056.07 |
| COLOR_LUMINACE_LAYOUT | 90.25 | 0.06 | 0.06 | 0.24 | 1.02 | 0.48 | 825.45 | 0.54 | 0.27 | 412.75 |
| COLOR_MPEG7_COLOR_LAYOUT | 7.52 | 0.04 | 0.04 | 0.21 | 0.87 | 0.31 | 505.31 | 0.46 | 0.18 | 252.68 |
| STRUCTURE_PROFILES | 1.02 | 0.02 | 0.02 | 0.13 | 0.84 | 0.11 | 636.57 | 0.43 | 0.06 | 318.29 |
| COLOREDGE_JCD | 53.55 | 0.03 | 0.03 | 0.16 | 0.75 | 0.29 | 805.58 | 0.39 | 0.16 | 402.80 |
| COLOREDGE_FCTH | 19.22 | 0.02 | 0.02 | 0.16 | 0.72 | 0.19 | 1,141.91 | 0.37 | 0.10 | 570.97 |
| COLOREDGE_CEDD | 10.78 | 0.02 | 0.02 | 0.15 | 0.65 | 0.22 | 691.12 | 0.34 | 0.12 | 345.57 |
| EDGE_MPEG7_EDGE_HISTOGRAM | 7.33 | 0.02 | 0.02 | 0.14 | 0.50 | 0.48 | 431.90 | 0.26 | 0.25 | 215.96 |
| EDGE_EDGEHIST | 41.04 | 0.01 | 0.01 | 0.11 | 0.25 | 0.22 | 147.39 | 0.13 | 0.12 | 73.70 |
| COLOR_THUMBNAIL | 6.47 | 0.00 | 0.00 | 0.07 | 0.23 | 1.25 | 34.61 | 0.12 | 0.63 | 17.31 |

Fig. 5: Quality metrics for the BSRN dataset. Generally, the Calinski Harabasz cluster quality index shows similar results to the Silhouette index.

### A. Quality Metric Evaluation

Figure 5 presents the impact of alternative parameter choices for our quality computation. We derive the data from the dataset showcased in Section V-B. For better interpretability, we depict the quality score ranking of our 24 feature descriptors in a scatterplot (Figure 6). The x-axis depicts the silhouette quality score $SC$ (the higher, the better), and the y-axis shows the feature descriptor's variance (the higher, the better). As one can see, the best performing feature descriptors are COLOR_OPPONENT_HISTOGRAM, its variation COLOR_-FUZZY_COLOR_OPPONENT_HISTOGRAM, and the COLOR_-AUTO_COLOR_CORRELOGRAM being present in the upper right quadrant. The COLOR_OPPONENT_HISTOGRAM descriptors use normalized color differences between all primary colors. Although the COLOR_OPPONENT_HISTOGRAM descriptor has not the highest variance scores, indicating feature discriminability, it outperforms the other feature descriptors. Our silhouette quality score shows a gradual ranking improvement without big jumps. On the other hand, we can also see—and validate from other experiments—that the Calinski Harabasz quality score shows some similarity with the ranking derived from the Silhouette score, while the Dunn index related score (needs to be considered inverse) shows a significant divergence. Generally, we found that both, the silhouette score and Calinski Harabasz score, often rank the same feature descriptor on the first place.

Note that this quality assessment is only valid and transferable between datasets if the underlying data representations expose similar visual structures. We also found that the feature ranking was entirely different for other datasets in our experiments. The appendix shows further experiments to validate this finding.

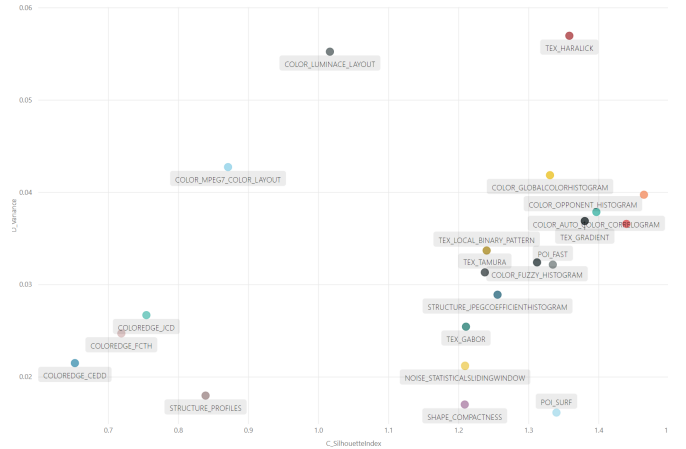

Fig. 6: The scatter plot representation of the feature descriptors performance. The y-axis shows the variance component of our quality score (more is better); the x-axis depicts the silhouette score (more is better). Nine feature descriptors result in similar quality metric scores. The texture descriptor "Haralick" scores first wrt. the variance component, but only fifth for the silhouette index, thus leading to an aggregate fourth QM score ranking place.

### B. Case Study: Earth Observation Data

We apply our exploration pipeline in a case study on scientific data from earth observation research. Our dataset contains a subset of the Baseline Surface Radiation Network

(BSRN) repository, maintained by the World Climate Research Programme [39]. The repository hosts data on measurements of water, sediment, ice and atmosphere, among others.

In our exploratory analysis, our primary goal is to develop an overview of the patterns contained in this dataset. Our quality metric evaluation, described in Section V-A, found that the COLOR_OPPONENT_HISTOGRAM differentiates patterns best. After exploring the first few Dendrogram Cutlevels (DC), we can assume that the descriptor differentiates mostly based on color density in the plot (Figure 4 shows DC4). After a closer inspection of the higher DC levels, however, we can also see that the feature descriptor seems to be able to group similar shapes while being rotation-invariant. Figure 7 shows some examples of the clustered patterns.

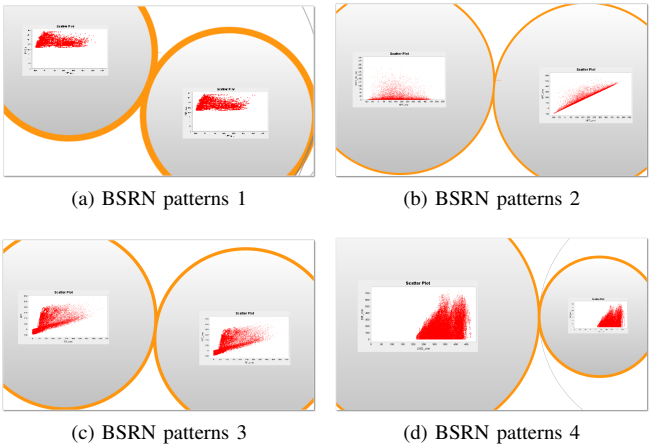

Fig. 7: Exemplary patterns of the BSRN dataset.

Generally, we found that scatter plots often share similar visual patterns if one of the axis dimensions remains static, such as "DIF" (Diffuse radiation) in Figure 7c or "LWD" (Long-wave downward radiation) in Figure 7d. While this finding seems obvious, we could use it as a starting point for finding deviations from this norm. As one example, we found that the pattern variability of scatter plots containing "DIR" (Direct radiation) seem to be worth examining.

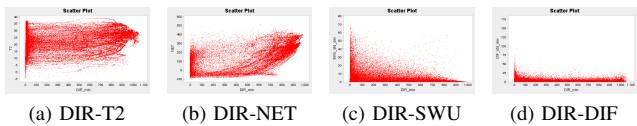

(a) DIR-T2   (b) DIR-NET   (c) DIR-SWU   (d) DIR-DIF

Fig. 8: Pattern variability of measure "DIR" (Direct radiation).

### C. Case Study: Genome Interaction Matrices

In a second case study, we are studying two collections of local patterns derived from two genome interaction matrices [33]. Genome interaction matrices capture pairwise interactions of up to 3 million regions on the genome and express several nested visual patterns, which act as a proxy to the spatial organization of the genome. Biologists study this spatial organization [11] as it has been shown to influence gene regulation [17], cell development [5], and pathogenic processes [20], [28], [41], [47]. In this case study we are focusing on two pattern types called loops and topologically associating domains (TADs) that ideally exhibit an off-diagonal pronounced peak or dot and an on-diagonal block respectively. The patterns have been detected previously [33] but their quality is very diverse [15], [24], [26] and generally not measurable due to a lack of ground truth.

The focus of our study is two-fold. First, we want to evaluate the overall quality of the pattern collection. And second, we try to stratify the group of patterns into subgroup showing potentially distinct biological events. With our quality metric evaluation we found that the NOISE_STATISTICALSLIDING-WINDOW feature descriptor differentiates patterns best [1]. The feature descriptor derives for subsequent regions in the image statistical information about color intensities. These sliding window values can be interpreted as a time series of color differences. The final NOISE_STATISTICALSLIDINGWINDOW feature vector describes the time series with respect to its average, variance, and standard deviation. We show the performance results of all other feature descriptors in our quality metric performance evaluation in the appendix. We start by exploring loop patterns globally within the GM12878 dataset from Rao et al. [33]. An ideal loop pattern is shown in Figure 9.a. After examining the first few DC of several chromosomes we realize that the diagonal is dominating the signal but starting with DC3 or DC4 we are able to identify clusters showing a pronounced loop pattern on average (Figure 9. highlighting that the NOISE_STATISTICALSLIDINGWINDOW descriptor is able to discriminate between noise and true dot-like pattern types, which we can use for to separate signal from noise.

## VI. DISCUSSION AND CONCLUSION

In this work, we present conceptually a pipeline for supporting a pattern-driven exploration of Big Data. As its foundation, we rely on an image feature analysis for given data representations and clustering to partition the pattern space into interpretable, coherent sub-clusters of patterns.

Yet, the underlying hypothesis is that at least one feature descriptor exists that is useful to quantify the existence of visual patterns in the dataset. The quantification of patterns in visualizations is an active research field with broadly two distinctive approaches: Either pattern measures are computed from the data space or the image space. Image-based quality metrics have the advantage that a direct correspondence to the human perceptual system is imminent. Following this argumentation, we can also assess the limits of our approach: an image-based pattern analysis can only work if the pattern space is clearly defined and distinguishable; i.e., the patterns can be discerned computationally and perceptually (human). Another limitation results from the choice of our data analysis machinery. Our quality scores rely on internal quality measures derived from feature vector distances. As, for example, shown by Reeb et al. [34] the choice of the dissimilarity calculation has an impact on the to be expected result interpretation. More research in this direction needs to be devoted to developing (more) dissimilarity score agnostic approaches.

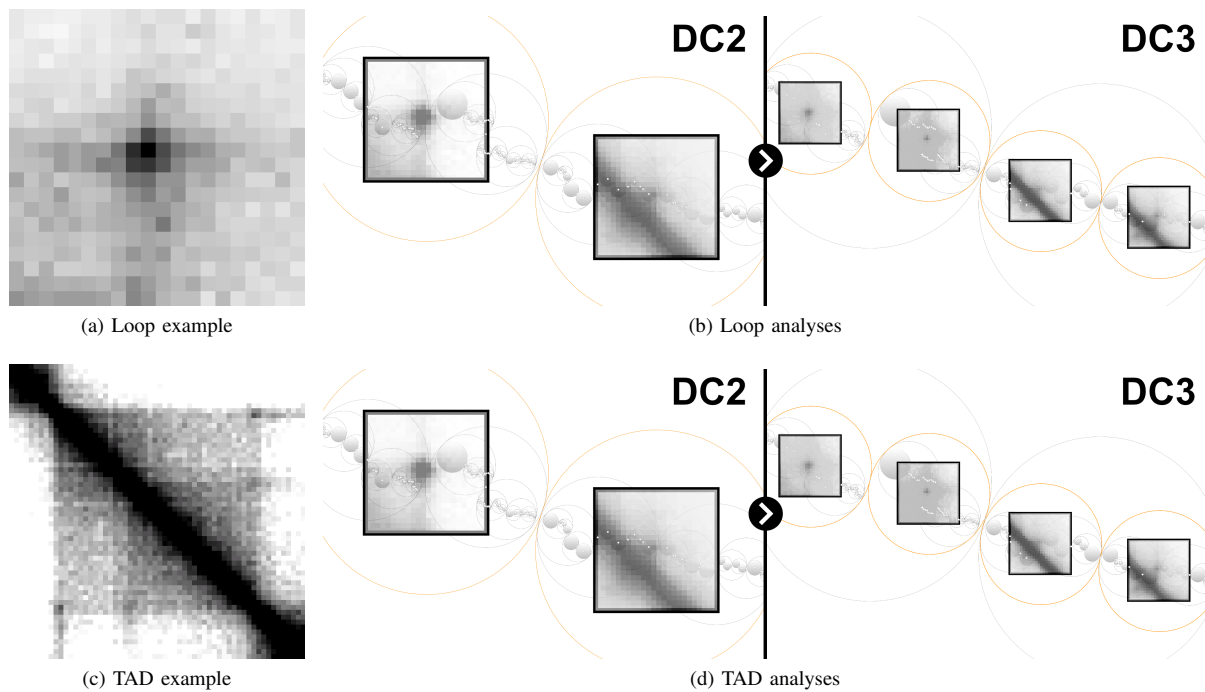

Fig. 9: Exemplary loop (a & b) and TAD (c & d) patterns of genome interaction matrices.

In general, we found that a pattern-driven analysis is a viable approach to guide the analysis of big datasets towards patterns with high support but also draws attention to the unexpected and outlying structures.

ACKNOWLEDGMENT

This work was supported in part by the National Institutes of Health (U01 CA200059 and R00 HG007583). We also thank Lin Shao for his valuable help in preparing the BSRN dataset.

REFERENCES

[1] G. Albuquerque, M. Eisemann, D. J. Lehmann, H. Theisel, and M. Magnor. Improving the visual analysis of high-dimensional datasets using quality measures. In *IEEE Conference on Visual Analytics Science and Technology*, pages 19–26. IEEE, 2010.
[2] B. Bach, N. Henry-Riche, T. Dwyer, T. Madhyastha, J.-D. Fekete, and T. Grabowski. Small multipiles: Piling time to explore temporal patterns in dynamic networks. In *Computer Graphics Forum*, volume 34, pages 31–40. Wiley Online Library, 2015.
[3] M. Behrisch, B. Bach, M. Hund, M. Delz, L. von Rüden, J.-D. Fekete, and T. Scheck. Magnostics: Image-based Search of Interesting Matrix Views for Guided Network Exploration. *IEEE Transactions on Visualization and Computer Graphics*, 23(1):31–40, Oct. 2017.
[4] M. Behrisch, F. Korkmaz, L. Shao, and T. Schreck. Feedback-Driven Interactive Exploration of Large Multidimensional Data Supported by Visual Classifier. In *IEEE Conference on Visual Analytics Science and Technology*, pages 43–52. IEEE CS Press, Oct. 2014. Peer-reviewed full paper.
[5] B. Bonev, N. M. Cohen, Q. Szabo, L. Fritsch, G. L. Papadopoulos, Y. Lubling, X. Xu, X. Lv, J.-P. Hugnot, A. Tanay, et al. Multiscale 3d genome rewiring during mouse neural development. *Cell*, 171(3):557–572, 2017.
[6] S. P. Borgatti, A. Mehra, D. J. Brass, and G. Labianca. Network analysis in the social sciences. *science*, 323(5916):892–895, 2009.
[7] S. Bremm, T. von Landesberger, J. Bernard, and T. Schreck. Assisted Descriptor Selection Based on Visual Comparative Data Analysis. *Computer Graphics Forum*, 30(3):891–900, June 2011.
[8] T. Caliński and J. Harabasz. A dendrite method for cluster analysis. *Communications in Statistics-theory and Methods*, 3(1):1–27, 1974.
[9] T. N. Dang and L. Wilkinson. Scagexplorer: Exploring scatterplots by their scagnostics. In *IEEE Pacific Visualization Symposium (PacificVis)*, pages 73–80. IEEE, March 2014.
[10] A. Dasgupta and R. Kosara. Pargnostics: Screen-space metrics for parallel coordinates. *IEEE Transactions on Visualization and Computer Graphics*, 16(6):1017–1026, Nov 2010.
[11] J. Dekker, A. S. Belmont, M. Guttman, V. O. Leshyk, J. T. Lis, S. Lomvardas, L. A. Mirny, C. C. Oshea, P. J. Park, B. Ren, et al. The 4d nucleome project. *Nature*, 549(7671):219, 2017.
[12] O. J. Dunn and V. A. Clark. *Applied statistics: analysis of variance and regression*. John Wiley & Sons, Inc., 1986.
[13] M. B. Eisen, P. T. Spellman, P. O. Brown, and D. Botstein. Cluster analysis and display of genome wide expression patterns. *Proceedings of the National Academy of Sciences*, 95(25):14863–14868, 1998.
[14] L. Ferreira and D. B. Hitchcock. A comparison of hierarchical methods for clustering functional data. *Communications in Statistics-Simulation and Computation*, 38(9):1925–1949, 2009.
[15] M. Forcato, C. Nicoletti, K. Pal, C. M. Livi, F. Ferrari, and S. Bicciato. Comparison of computational methods for hi-c data analysis. *Nature methods*, 14(7):679, 2017.
[16] E. B. Fowlkes and C. L. Mallows. A method for comparing two hierarchical clusterings. *Journal of the American statistical association*, 78(383):553–569, 1983.
[17] P. Fraser and W. Bickmore. Nuclear organization of the genome and the potential for gene regulation. *Nature*, 447(7143):413–417, 2007.
[18] J. Friedman and J. Tukey. A projection pursuit algorithm for exploratory data analysis. *IEEE Transactions on Computers*, C-23(9):881–890, Sept 1974.
[19] F. D. Gibbons and F. P. Roth. Judging the quality of gene expression-based clustering methods using gene annotation. *Genome research*, 12(10):1574–1581, 2002.

[20] D. Hnisz, A. S. Weintraub, D. S. Day, A.-L. Valton, R. O. Bak, C. H. Li, J. Goldmann, B. R. Lajoie, Z. P. Fan, A. A. Sigova, J. Reddy, D. Borges-Rivera, T. I. Lee, R. Jaenisch, M. H. Porteus, J. Dekker, and R. A. Young. Activation of proto-oncogenes by disruption of chromosome neighborhoods. *Science*, 351(6280):1454–1458, 2016.

[21] P. K. Hopke and L. Kaufman. The use of sampling to cluster large data sets. *Chemometrics and Intelligent Laboratory Systems*, 8(2):195 – 204, 1990.

[22] L. Hubert and P. Arabie. Comparing partitions. *Journal of Classification*, 2(1):193–218, Dec 1985.

[23] P. Kerpedjiev, N. Abdennur, F. Lekschas, C. McCallum, K. Dinkla, H. Strobelt, J. M. Luber, S. B. Ouellette, A. Ahzir, N. Kumar, J. Hwang, B. H. Alver, H. Pfister, L. A. Mirny, P. J. Park, and N. Gehlenborg. Higlass: Web-based visual comparison and exploration of genome interaction maps. *bioRxiv*, 2017.

[24] P. Kerpedjiev, N. Abdennur, F. Lekschas, C. McCallum, K. Dinkla, H. Strobelt, J. M. Luber, S. B. Ouellette, A. Azhir, N. Kumar, et al. Higlass: Web-based visual exploration and analysis of genome interaction maps. *bioRxiv*, page 121889, 2018.

[25] D. J. Lehmann, F. Kemmler, T. Zhyhalava, M. Kirschke, and H. Theisel. Visualnostics: Visual guidance pictograms for analyzing projections of high-dimensional data. *Computer Graphics Forum*, 34(3):291–300, 2015.

[26] F. Lekschas, B. Bach, P. Kerpedjiev, N. Gehlenborg, and H. Pfister. Hipiler: Visual exploration of large genome interaction matrices with interactive small multiples. *IEEE transactions on visualization and computer graphics*, 24(1):522–531, 2018.

[27] A. Lex, M. Streit, H.-J. Schulz, C. Partl, D. Schmalstieg, P. J. Park, and N. Gehlenborg. Stratomex: Visual analysis of large-scale heterogeneous genomics data for cancer subtype characterization. In *Computer graphics forum*, volume 31, pages 1175–1184. Wiley Online Library, 2012.

[28] D. G. Lupiáñez, K. Kraft, V. Heinrich, P. Krawitz, F. Brancati, E. Klopocki, D. Horn, H. Kayserili, J. M. Opitz, R. Laxova, et al. Disruptions of topological chromatin domains cause pathogenic rewiring of gene-enhancer interactions. *Cell*, 161(5):1012–1025, 2015.

[29] J. Mackinlay. Automating the design of graphical presentations of relational information. *ACM Transactions On Graphics (TOG)*, 5(2):110–141, 1986.

[30] G. W. Milligan and M. C. Cooper. A study of the comparability of external criteria for hierarchical cluster analysis. *Multivariate behavioral research*, 21 4:441–58, 1986.

[31] S. M. Paley and P. D. Karp. The pathway tools cellular overview diagram and omics viewer. *Nucleic Acids Research*, 34(13):3771–3778, 2006.

[32] W. M. Rand. Objective criteria for the evaluation of clustering methods. *Journal of the American Statistical association*, 66(336):846–850, 1971.

[33] S. S. Rao, M. H. Huntley, N. C. Durand, E. K. Stamenova, I. D. Bochkov, J. T. Robinson, A. L. Sanborn, I. Machol, A. D. Omer, E. S. Lander, et al. A 3d map of the human genome at kilobase resolution reveals principles of chromatin looping. *Cell*, 159(7):1665–1680, 2014.

[34] P. D. Reeb, S. J. Bramardi, and J. P. Steibel. Assessing dissimilarity measures for sample-based hierarchical clustering of rna sequencing data using plasmode datasets. *PLOS ONE*, 10(7):1–18, 07 2015.

[35] M. Z. Rodriguez, C. H. Comin, D. Casanova, O. M. Bruno, D. R. Amancio, F. A. Rodrigues, and L. d. F. Costa. Clustering algorithms: A comparative approach. *arXiv preprint arXiv:1612.08388*, 2016.

[36] L. Rokach and O. Maimon. Clustering methods. In *Data mining and knowledge discovery handbook*, pages 321–352. Springer, 2005.

[37] P. J. Rousseeuw. Silhouettes: a graphical aid to the interpretation and validation of cluster analysis. *Journal of computational and applied mathematics*, 20:53–65, 1987.

[38] Y. Rui, T. S. Huang, and S.-F. Chang. Image retrieval: Current techniques, promising directions, and open issues. *Journal of visual communication and image representation*, 10(1):39–62, 1999.

[39] M. Scherer, T. v. Landesberger, and T. Schreck. A benchmark for content-based retrieval in bivariate data collections. In *Conference on Theory and Practice of Digital Libraries*, pages 286–297, 2012.

[40] J. Schneidewind, M. Sips, and D. A. Keim. Pixnostics: Towards measuring the value of visualization. In *IEEE Symposium on Visual Analytics Science and Technology*, pages 199–206. IEEE, Oct 2006.

[41] L. Seaman, H. Chen, M. Brown, D. Wangsa, G. Patterson, J. Camps, G. S. Omenn, T. Ried, and I. Rajapakse. Nucleome analysis reveals structure–function relationships for colon cancer. *Molecular Cancer Research*, 15(7):821–830, 2017.

[42] L. Shao, T. Schleicher, M. Behrisch, T. Schreck, I. Sipiran, and D. Keim. Guiding the exploration of scatter plot data using motif-based interest measures. In *IEEE Symposium on Big Data Visual Analytics*, 2015.

[43] L. Shao, T. Schleicher, M. Behrisch, T. Schreck, I. Sipiran, and D. A. Keim. Guiding the exploration of scatter plot data using motif-based interest measures. *J. Vis. Lang. Comput.*, 36:1–12, 2016.

[44] M. Sips, P. Kthur, A. Unger, H.-C. Hege, and D. Dransch. A Visual Analytics Approach to Multiscale Exploration of Environmental Time Series. *IEEE Transactions on Visualization and Computer Graphics*, 18(12):2899–2907, Dec. 2012.

[45] M. Sips, B. Neubert, J. P. Lewis, and P. Hanrahan. Selecting good views of high-dimensional data using class consistency. *Computer Graphics Forum*, 28(3):831–838, 2009.

[46] A. W. Smeulders, M. Worring, S. Santini, A. Gupta, and R. Jain. Content-based image retrieval at the end of the early years. *IEEE Transactions on pattern analysis and machine intelligence*, 22(12):1349–1380, 2000.

[47] M. Spielmann, D. G. Lupiáñez, and S. Mundlos. Structural variation in the 3d genome. *Nature Reviews Genetics*, 2018.

[48] H. Strobelt, E. Bertini, J. Braun, O. Deussen, U. Groth, T. Mayer, and D. Merhof. Hitsee knime: a visualization tool for hit selection and analysis in high-throughput screening experiments for the knime platform. *BMC Bioinformatics*, 13(Suppl 8):S4, Dec. 2012.

[49] A. Tatu, G. Albuquerque, M. Eisemann, P. Bak, H. Theisel, M. A. Magnor, and D. A. Keim. Automated analytical methods to support visual exploration of high-dimensional data. *IEEE Transactions on Visualization and Computer Graphics*, 17(5):584–597, May 2011.

[50] A. Tatu, F. Maass, I. Faerber, E. Bertini, T. Schreck, T. Seidl, and D. A. Keim. Subspace Search and Visualization to Make Sense of Alternative Clusterings in High-Dimensional Data. In *IEEE Conference on Visual Analytics Science and Technology*, pages 63–72. IEEE, IEEE CS Press, Oct 2012.

[51] J. W. Tukey and P. A. Tukey. Computer graphics and exploratory data analysis: An introduction. *The Collected Works of John W. Tukey: Graphics: 1965-1985*, 5:419, 1988.

[52] M. O. Ward and Z. Guo. Visual exploration of time-series data with shape space projections. *Computer Graphics Forum*, 30(3):701–710, 2011.

[53] L. Wilkinson, A. Anand, and R. Grossman. Graph-theoretic scagnostics. In *IEEE Symposium on Information Visualization*, volume 5, pages 157–164. IEEE Computer Society, Oct 2005.

[54] J. Wood, A. Slingsby, and J. Dykes. Visualizing the dynamics of London's bicycle-hire scheme. *Cartographica: The International Journal for Geographic Information and Geovisualization*, 46(4):239–251, 2011.

[55] J. Yang, D. Hubball, M. O. Ward, E. A. Rundensteiner, and W. Ribarsky. Value and relation display: Interactive visual exploration of large data sets with hundreds of dimensions. *IEEE Transactions on Visualization and Computer Graphics*, 13(3):494–507, 2007.